\documentclass{kluwer}    % Specifies the document style.
\usepackage{psfig}
\begin{document}                                                               
\begin{article}
\begin{opening}         
\title{NGC 4340: Double Bar + Fossil Nuclear Ring} 
\author{Peter \surname{Erwin}}
\author{Juan Carlos \surname{Vega Beltr\'an}}  
\author{John \surname{Beckman}}  
\runningauthor{Vega Beltr\'an {\it et al.}}
\runningtitle{NGC4340: Double Bar + Fossil Nuclear Ring}
\institute{IAC, Tenerife, Spain}
%\date{April 15, 1993}

%\begin{abstract}
%This is a sample input file.  Comparing it with the output it
%generates can show you how to produce a simple document of
%your own.
%\end{abstract}
\end{opening}           

\begin{figure} 
\includegraphics{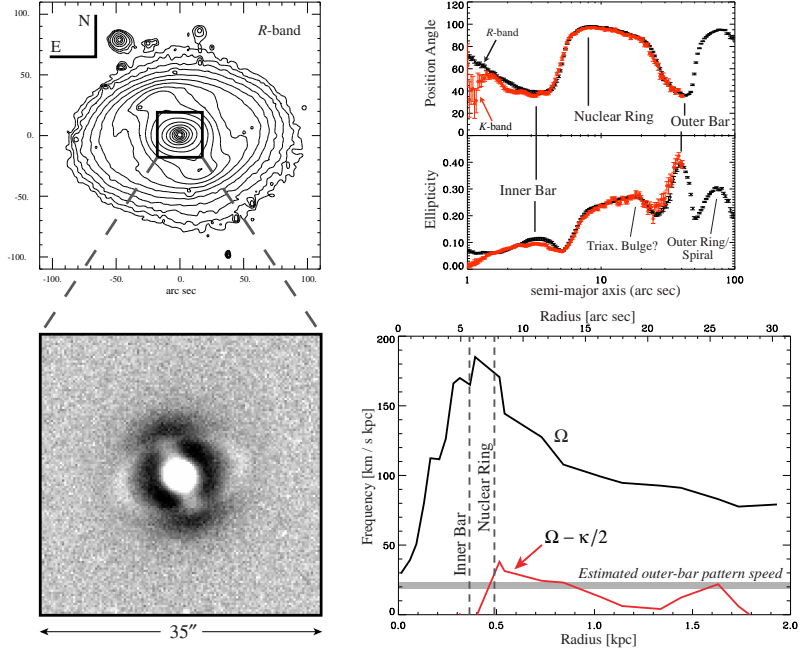} 
\vspace*{-1.4cm}

\caption[]{NGC 4340 is a double-barred SB0 galaxy in the Virgo cluster
(Wozniak \textit{et al.} 1995, A\&AS 111: 115).  Here, we show that
this galaxy also posseses a luminous \textit{stellar} nuclear ring of
relatively old stars with little or no gas.  The ring lies just
outside the inner bar, at the probable inner inner Lindblad resonance
(IILR) of the outer bar.  Careful inspection of the isophotes and
unsharp masks shows that the two bars are slightly misaligned,
which suggests they may be independently rotating.
\vspace*{0.25cm}\\

The bright nuclear ring distorts the isophotes and ellipse fits (upper
right: black are $R$-band, grey are $K$-band) and is most clearly seen
in the unsharp mask (lower left; inner bar also visible).  $B\!-\!R$
color maps show no features associated with the ring: it is the same
color as the surrounding bulge, and thus probably an old, ``fossil''
remnant of an earlier star-formation episode.  We use Simien \&
Prugniel's (1997, A\&AS 126: 15) major-axis velocities to compute
resonance curves (lower right); we use our spectrum along the
outer-bar major axis (not shown) to estimate that bar's pattern speed.
The $\Omega - \kappa/2$ curve shows that any ILRs which may be present
should lie at $r > 7$ arc sec.  This places the inner bar within the
(inner) ILR and suggests that the nuclear ring is at or just inside
the same ILR; this agrees with theoretical arguments (Pfenniger \&
Norman 1990, ApJ 363: 391; Maciejewski \& Sparke 2000, MNRAS 313:
745).}

\label{NGC4340ring}
\end{figure}

\end{article}
\end{document}